The oligopoly of academic publishers persists in exclusive database

Simon van Bellen[1], Juan Pablo Alperin[2,3], Vincent Larivière[1,4,5,6]

[1] Consortium Érudit, Montréal, QC, Canada

[2] School of Publishing, Simon Fraser University, Canada

[3] Public Knowledge Project, Simon Fraser University, Canada

[4] École de bibliothéconomie et des sciences de l'information, Université de Montréal, Montréal, QC, Canada

[5] Observatoire des Sciences et des Technologies, Centre interuniversitaire de recherche sur la science et la technologie, Université du Québec à Montréal, Montréal, QC, Canada.

[6] Department of Science and Innovation-National Research Foundation Centre of Excellence in Scientometrics and Science, Technology and Innovation Policy, Stellenbosch University, Stellenbosch, Western Cape, South Africa.

## Abstract

Global scholarly publishing has been dominated by a small number of publishers for several decades. We aimed to revisit the debate on corporate control of scholarly publishing by analyzing the relative shares of major publishers and smaller, independent publishers. Using the Web of Science, Dimensions and OpenAlex, we managed to retrieve twice as many articles indexed in Dimensions and OpenAlex, compared to the rather selective Web of Science. As a result of excluding smaller publishers, the 'oligopoly' of scholarly publishers persists, at least in appearance, according to the Web of Science. However, both Dimensions' and OpenAlex' inclusive indexing revealed the share of smaller publishers has been growing rapidly, especially since the onset of large-scale online publishing around 2000, resulting in a current cumulative dominance of smaller publishers. While the expansion of small publishers was most pronounced in the social sciences and humanities, the natural and medical sciences showed a similar trend. A major geographical divergence is also revealed, with some countries, mostly Anglo-Saxon and/or located in northwestern Europe, relying heavily on major publishers for the dissemination of their research, while others being relatively independent of the oligopoly, such as those in Latin America, northern Africa, eastern Europe and parts of Asia. The emergence of digital publishing, the reduction of expenses for printing and distribution and open-source journal management tools may have contributed to the emergence of small publishers, while the development of inclusive bibliometric databases has allowed for the effective indexing of journals and articles. We conclude that enhanced visibility to recently created, independent journals may favour their growth and stimulate global scholarly bibliodiversity.

1. Introduction

Global scholarly publishing has been dominated by a restricted number of publishers since at least the 1970s (1). The market share of the world's largest publishers, which include RELX (Elsevier), Springer Nature, Wiley, Taylor & Francis, and Sage Publications, in the global production of scholarly journals and articles has grown continually since the 1970s (2). Using the Web of Science (WoS), Larivière *et al*. (2015) quantified the growing share of research articles controlled by those publishers, in terms of numbers of scholarly journals and articles, the latter attaining more than 50% in 2013 (1). They also showed how those proportions varied across disciplines, with the social sciences being the most highly concentrated (70%) and the humanities remaining relatively independent (20%). The majority of papers in medical and natural sciences, where most research papers are published and where journals are the most expensive (3), were controlled by the major five publishers. The dominance of the major publishing companies accelerated during the 1990s (1), coinciding with the onset of large-scale digital publishing.

In the last decade, the scholarly publishing landscape has continued to evolve, especially as open access (OA) has become mainstream (4). Growth in OA mirrored an increasing reliance on article processing charges (APCs) by both the traditional oligopoly (Butler et al., 2023) and new APC-only publishers entering the market (Hanson et al., 2024). Yet many independent OA journals adopted new, open-source publishing software, such as Open Journal Systems (OJS), which has streamlined journal management and facilitated the creation of digital journals, especially outside North America and Europe (5). Nevertheless, both developments also facilitated the rise of questionable (so-called 'predatory') journals, i.e., those that do not adhere to ethical publication practices (6,7).

Recent works have assessed the relative importance of corporate publishers in the research dissemination system, mostly in the context of OA publishing models. Zhang et al. (8) have shown how publishers' business models are changing, that their revenues are increasingly relying on Article Processing Charges (APCs) rather than subscriptions, and that this changing model is driving the recent mergers and acquisitions observed in the industry. This can be exemplified by the recent acquisition of Hindawi by Wiley, or by the investments made by the Holtzbrinck Publishing Group, now the major shareholder of Springer Nature, in Frontiers media (9). Drawing on journal's APC price lists for the top five most prolific publishers in terms of articles (RELX, Springer Nature, Wiley, Taylor & Francis and Sage Publications), Butler et al. (10) quantified the revenue stream coming from APCs, which exceeded more than 1 billion $US for 2015–2018. The majority of those revenues were obtained through Gold OA journals, although RELX and Springer Nature obtained the majority of their OA revenues from hybrid journals.

In parallel to changes in the scholarly publishing industry, the last decades have seen significant progress in open-source practices and applications, driven by both technology and a socio-cultural willingness among scholars to render new knowledge more accessible (11). One exemplification of this trend is the development of inclusive bibliometric databases, such as Dimensions (12,13) and OpenAlex (14). These databases rely on increasingly powerful algorithms, the growth in use of persistent identifiers and the openness of article metadata, mostly through Crossref. Both Dimensions and OpenAlex follow a similar approach to indexing and have been shown to have similar coverage (15–17). Dimensions is freely available only for research purposes, while OpenAlex is a not-for-profit initiative and makes its data and code fully open.

These recently developed databases offer an opportunity to re-assess the extent to which major publishers control the research dissemination landscape without the limitations of incomplete coverage offered by more selective bibliographic databases, such as WoS and Scopus. Applying selective criteria for inclusion, based on their citedness, WoS and Scopus purposely index a restricted set of journals which represents a small percentage of all scholarly journals published worldwide, as suggested by comparisons with more inclusive databases such as Dimensions (18) or Ulrichsweb (19). Both WoS and Scopus are biased towards the natural sciences and medicine, English-language content (19,20) and journals and publishers that are outside North America and Europe. Shu and Larivière (9) used restrictive (WoS) and inclusive (Dimensions) databases to assess the level of concentration in the OA publishing market. They showed how that market is increasingly concentrated in WoS –which suggests concentration in the elite set of journals– while it is decreasing for most of the period in Dimensions.

The aim of this article is to revisit the debate on corporate control of scholarly publishing, by comparing restrictive and inclusive data sources, and to assess how specific countries' papers are affected by the concentration of the publishing market.

## 2. Materials and Methods

Data was retrieved from WoS, Dimensions and OpenAlex databases. Access to WoS was obtained through the OST server at Université du Québec à Montréal. Dimensions was accessed using BigQuery and OpenAlex' data was retrieved through the platform developed by the Curtin Open Knowledge Initiative (COKI). We restricted our analyses to works that have been published in journals with an ISSN and that have a DOI assigned (Figure 1). These criteria allowed for an optimal comparison with the indexing of WoS, which is mostly restricting its coverage to journal articles.

Information on journals' publishers—and especially on their ownership structures—often remains incomplete in bibliometric databases. Therefore, we had to perform significant data cleaning and reorganization to take account of mergers and acquisitions, as well as of the numerous divisions belonging to the same publisher. In Dimensions and OpenAlex, some content was associated with aggregators, rather than publishers. As we focused on publishers' concentrations, these entities, which included JSTOR, SciELO and CAIRN, were not included in the rankings of the degree of concentration, but the articles disseminated by these were considered in calculating the publishers' degree of concentration. The content of some questionable publishers was removed from the datasets, based on an updated version of Beall's list. These included OMICS, Sophia Publishing Group and Academic Journals, but some questionable content is likely to have been included as the evaluation could not be performed in high detail.

As the publishing landscapes of natural sciences and medicine, social sciences, and humanities exhibit important differences in terms of corporate control (1), dissemination languages (21) and open access (4), among others, we made distinctions in output according to discipline. All three databases use different field classifications, which are in some cases relying on journals (WoS) and in others relying on automatic classification of papers through algorithms (Dimensions and OpenAlex). In Dimensions, articles may be assigned to multiple fields of research. In order to correctly classify articles between Natural and Medical Sciences, Social Sciences and Humanities, only articles that had a single field assigned were retained for the analyses.

Figure 1: Article indexing in a) Dimensions and b) OpenAlex according to their classification as being a journal article, the presence of an ISSN for the journal, and the presence of a DOI for the article, covering the 1980-2021 period.

## 3. Results

**3.1 Comparing bibliometric databases**

The number of scholarly articles and journals published between 1980 and 2021 is remarkably higher according to Dimensions and OpenAlex, compared to WoS. Whereas WoS returned around 43 million articles in total, Dimensions and OpenAlex each contained around 80 million articles. Differences in the numbers of journals are even more striking. While both Dimensions and OpenAlex show an accelerating growth in journals since 2000, WoS suggests only a slight increase during the last 25 years (Figure 2). In terms of articles, all databases return increasing numbers since 1980, but growth appears much quicker according to Dimensions and OpenAlex. This suggests that the growth of WoS has not followed that of scholarship (22) and that this indexing gap is increasing.

Figure 2: Trends in the numbers of articles and journal titles published annually, according to three bibliometric databases, 1980-2021.

Dimensions and WoS have a greater overlap in journal collections than OpenAlex and WoS. On the other hand, OpenAlex has a greater number of journals that are lacking from both Dimensions and WoS (Figure 3).

Figure 3: Comparison of WoS, Dimensions and OpenAlex journal collections, matched on ISSN (2019-2021).

## 3.2 The share of the major publishers

Major differences appear between the three databases when quantifying the relative share of articles for each publisher (Figure 4). Corroborating the findings by Larivière et al. (2015), WoS shows an increasing concentration of scholarly journal articles among the major publishers since the 1980s. However, when considering the expanded coverage of both Dimensions and OpenAlex, an opposite trend emerges, with decreasing shares of these publishers, especially since the mid-1990s. The three databases converge in terms of the relative stability of the *composition* of the group of major publishers: RELX (Elsevier), Springer Nature, Wiley, MDPI and Taylor & Francis are the current five largest publishers in terms of the volume of articles (Figure 4).

Specifically focusing on the share of these major five publishers, WoS shows an increasing concentration of articles with the major publishers increasing from 35% in 1980 to 59% in 2021. In contrast, both Dimensions and OpenAlex show that the five largest publishers own a decreasing share of indexed content during the last two decades. In Dimensions, this share went down from 54% in 1997 to 37% in 2021 and in OpenAlex it passed from 55% to 38% during the same period. Trends are similar for the top-20 publishers.

Figure 4: Proportion of articles published by a selection of major publishers, for those that have been part of the major five publishers at some point between 1980 and 2021, for each database.

The concentration of *journals* among the main publishers shows a massive decline during the last 25 years according to Dimensions and OpenAlex, both suggesting a mere 16% of journals are currently published by the five major publishers. Again, the WoS data contradicts this trend, with increasing concentration of both articles and journals since the 1980s. According to WoS, the share of articles published by the five biggest publishers has continued to rise since 2005, yet the share of journals has remained stable during this period.

These overall trends hold across disciplines but are less pronounced in the natural and medical sciences in comparison with the social sciences and the humanities (Figure 5). It also appears WoS and Dimensions diverge more significantly for the social sciences and the humanities. Interestingly, whereas the social sciences appear more concentrated than the natural sciences and medicine according to the WoS, Dimensions suggests the opposite.

Figure 5: Proportions of articles published by the main publishers, per major discipline, according to two bibliometric databases, 1980-2021.

The declining proportion of articles and journals during the 1990s, obtained using Dimensions and OpenAlex, is related to their indexing approaches. The indexing of Dimensions and OpenAlex are more inclusive than that of WoS, as they rely heavily on Crossref data. This explains the inclusion of vast numbers of smaller journals since the early 2000s, a majority of which may have been founded as digital and many of which may belong to the social sciences and the humanities. The relative absence of independent journals is what characterizes WoS. As a result, 50% of the journals in WoS are published by the major five publishers - in terms of number of journals published between 2019 and 2021—RELX, Springer Nature, Taylor & Francis, Wiley and Sage Publishing.

**3.3 Geographical divergence in the dominance of the main publishers**

In addition to the variations associated with disciplines, there is a strong geographic divergence in the use of commercial publishers' journals by researchers. Dimensions' data show that in the natural and medical sciences the vast majority of the countries publish more than half of their articles in journals hosted by the major five publishers (Figure 6). Nevertheless, several regions and countries show much lower proportions, in particular Indonesia (14%), Ukraine (27%), Russia (28%) and several central Asian and central American countries.

The relative use of commercial publishers' journals by researchers is more polarized in social sciences and humanities (Figure 6). In the social sciences, many countries show a very limited use of commercial publishers' journals for the publication of their research, accounting for less than 30% of the articles. These countries are mostly located in Latin America, eastern Europe, central Asia and parts of northern Africa, in addition to Indonesia (5%) and Russia (11%), forming what we may call a 'Noligopoly Belt', spanning across the globe. Still, western Europe, North America and China generally show values well exceeding 50%. This contrasting trend is similar in the humanities, yet the commercial publishers' share is still lower overall.

Figure 6: Proportion of articles published by the five major publishers (based on the number of articles), based on data from Dimensions (2019-2021). Articles were attributed to a country if at least one author originated from it.

4. Discussion

**4.1 WoS' authority and indexing practices**

The restrictive selection criteria of WoS reinforce the significance of the publishers in two interrelated ways. First and most simply, by ignoring the long tail of scholarly journals the WoS arbitrarily limits the denominator used to calculate the share of works which gives the appearance that those journals and publishers that are included are somehow more important. However, just as significantly, this apparent importance serves to drive additional articles to those journals and additional journals to those publishers. Moreover, the criteria for inclusion in WoS are more generally met by journals published by the main commercial publishers and may be difficult to fulfill by smaller, independent publishers, as they are based on citations (counted from within WoS), so-called "international impact", certain technical and editorial characteristics and a preference for the use of English.

**4.2 Digital publishing and the emergence of new journals**

The proliferation of smaller, independent journals in the 2000s may be linked to the development of the Web, drastically enhancing their potential readership. It may also be related to the abolishment of print and the reduction of the expenses related to distribution making scholarly publishing more cost-effective. The creation and persistence of smaller journals was facilitated by the development of open publishing platforms, such as Open Journal Systems (OJS), which was launched in 2002. Open infrastructures help to streamline manuscript handling, facilitate communication with authors and enhance discoverability and indexing. Tens of thousands of scholarly journals use OJS today (5). Of those active during the 2019-2021 period, a majority is indexed in Dimensions (56%) and OpenAlex (63%), yet only 1% of journals using OJS are indexed in WoS (Figure 7), confirming that WoS very poorly represents these independent journals - and global publishing in general.

Dimensions and OpenAlex effectively shed light on publications that, until very recently, could hardly be traced. Yet, even these tools cannot manage to capture all scientific publications as the use of persistent identifiers, such as journal (ISSN) and article (DOI), and the availability of appropriate metadata is limited (23). For example, only 73% of 120,232 active journals present in OpenAlex (2019-2021) are identifiable by an ISSN. Specific article metadata such as author affiliations are even more rare, especially among smaller publishers: according to Dimensions, in natural sciences and medicine, 73% of their articles (2019-2021) have the author's country of affiliation indexed, compared to 48% in social sciences and 39% in humanities. This implies that the proportions of the articles published at the major publishers as presented in Figure 6 are actually overestimated, especially for countries that already show low reliance on the major publishers.

Figure 7: Comparison of the journal indexing of Web of Science, Dimensions and OpenAlex (2019-2021) and journals using OJS, including journals not using DOI for articles. The relationships (horizontal axis) are ordered by the degree of overlap between sets.

## 4.3 The current state of the oligopoly

The oligopoly of corporate publishers persists at a global level, but its dominance is partial. It remains strong among the journals included in major, yet selective databases like the WoS. It also prevails in Western Europe, North America and China, and especially in the natural sciences and medicine. In parallel, however, the global major publishers are much less dominant for many countries in the global South, such as Brazil, Indonesia and Russia. These important differences between countries may have multiple causes. In some cases, countries have strongly centralized scientific policies and well-established support for domestic, mainly not-for-profit journals, which is the case in both Indonesia and Brazil. On the other hand, strong national incentives to publish in 'high-impact', international journals may be as frequent, for example in China (24), which only recently abandoned direct financial rewards for its researchers to this effect, but also in some western countries (25,26). The relative dominance of major commercial publishers appears 'natural' in countries having English as a national language, or having a strong Anglo-Saxon influence, such as Australia, New Zealand, South Africa, Kenya, Tanzania and Ghana. A long history of pursuing research in national (non-English) languages may make countries refrain from publishing among commercial publishers. This may be the case for Russia and Eastern Europe.

Although the position of the traditional major publishers, such as Elsevier, Springer Nature, Wiley and Taylor & Francis, may appear weakened in terms of sheer publishing numbers, they still represent a major share of the most visible research, managing to perpetuate the symbolic capital associated with their journals. Yet, in the context of digital publishing, we may consider questioning the relevance of the traditional dissemination and archiving functions of the journal itself (27), as in the digital era, access to articles generally bypasses the journal's cover page. In addition, and in line with the philosophy behind the DORA and CoARA initiatives, we should question the relevance of the journal *notoriety* in estimating article 'quality' or 'impact'. Valorizing the major commercial publishers' journals likely contributes to the relative marginalization of smaller, independent journals.

The dominance of the major publishers may nowadays be oriented differently. For more than a decade, they have aimed to diversify their products and they will likely continue to do so. For example, RELX (Elsevier) has a long record of developing applications for management and showcasing of research and publications, information systems for tracking faculty accomplishments, reference software, monitors for research performance and scientific document text mining applications. Commercial publishers have focused on artificial intelligence applications for some period, with a growing interest in building tools powered by Large Language Models trained on, at least partly, their own corpus.

## 5. Conclusions

Globally, the major scientific publishers may have lost their dominance in terms of journal and article volumes, mainly due to the explosive growth in journals and articles published elsewhere. Mirroring the development of the Web since the late 1990s, the emergence of online publishing is likely one of the main factors contributing to this trend. The development of open publishing platforms such as OJS facilitated the creation of journals. The potential of literature findability and accessibility has increased markedly, which may have been to the advantage of independent journals. In addition, the creation and the development of inclusive databases, such as Dimensions and OpenAlex, has enabled us to identify and study the emergence of independent publishers. Providing such visibility to recently created, independent journals may favour their growth and contribute to enhancing the global diversity in scholarly publishing.

## 6. Acknowledgments

We thank Cameron Neylon (Curtin University) for providing access to the COKI tool and assistance in usage.

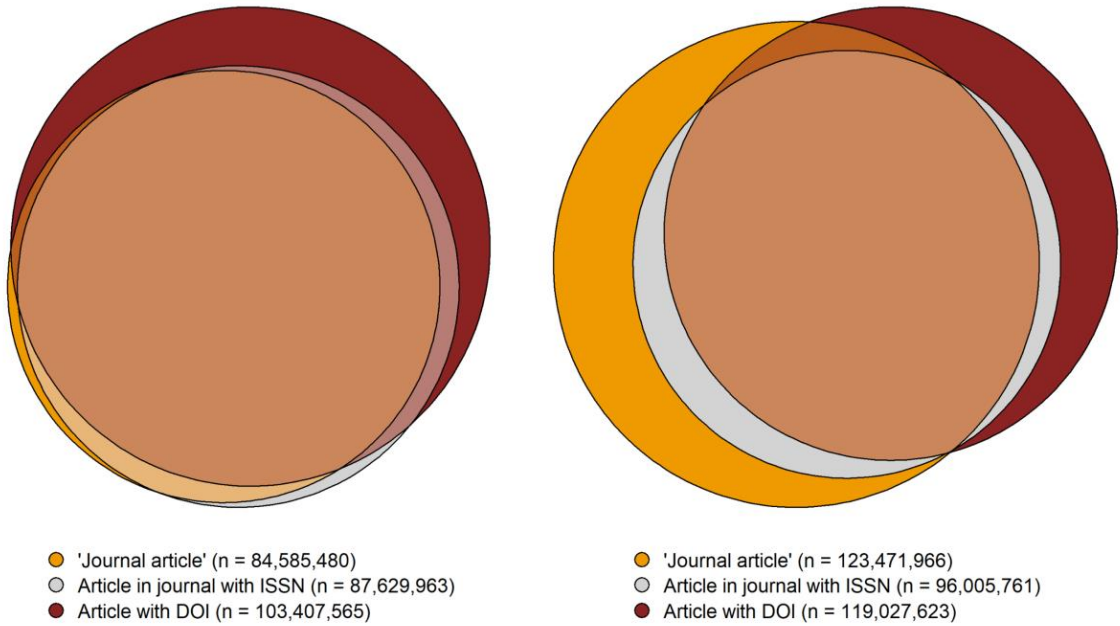

Figure 1: Article indexing in a) Dimensions and b) OpenAlex according to their classification as being a journal article, the presence of an ISSN for the journal, and the presence of a DOI for the article, covering the 1980-2021 period.

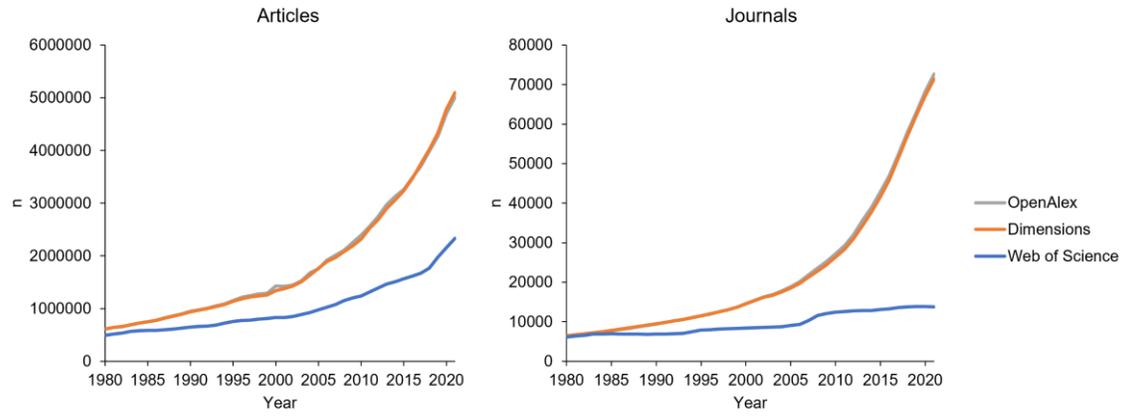

Figure 2: Trends in the numbers of articles and journal titles published annually, according to three bibliometric databases, 1980-2021.

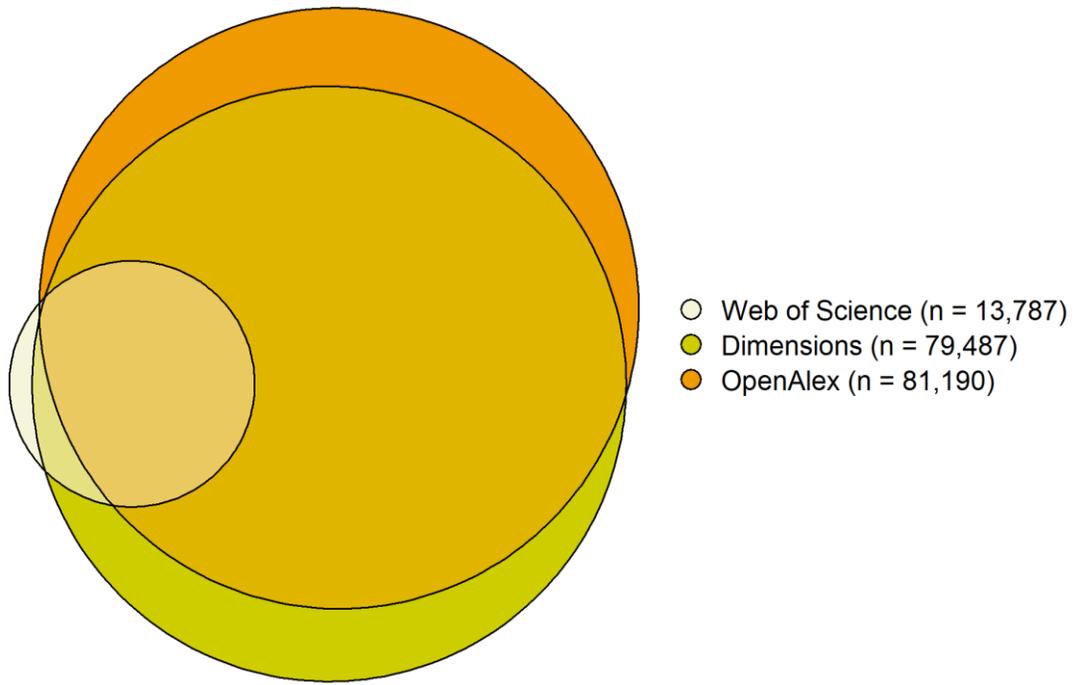

Figure 3: Comparison of WoS, Dimensions and OpenAlex journal collections, matched on ISSN (2019-2021).

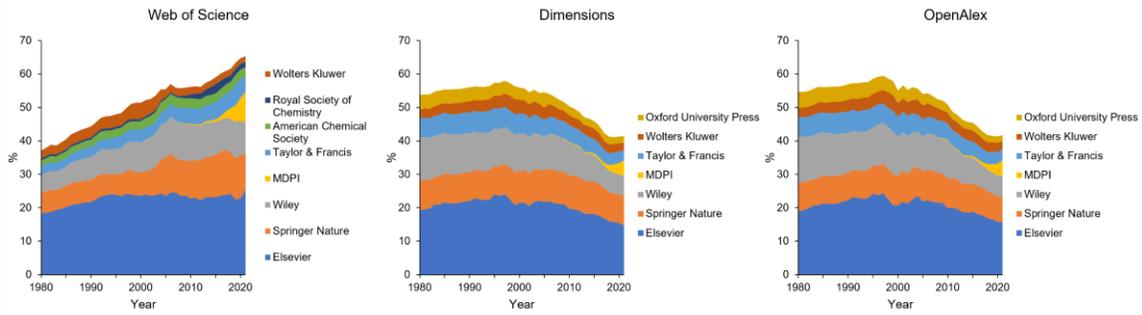

Figure 4: Proportion of articles published by a selection of major publishers, for those that have been part of the major five publishers at some point between 1980 and 2021, for each database.

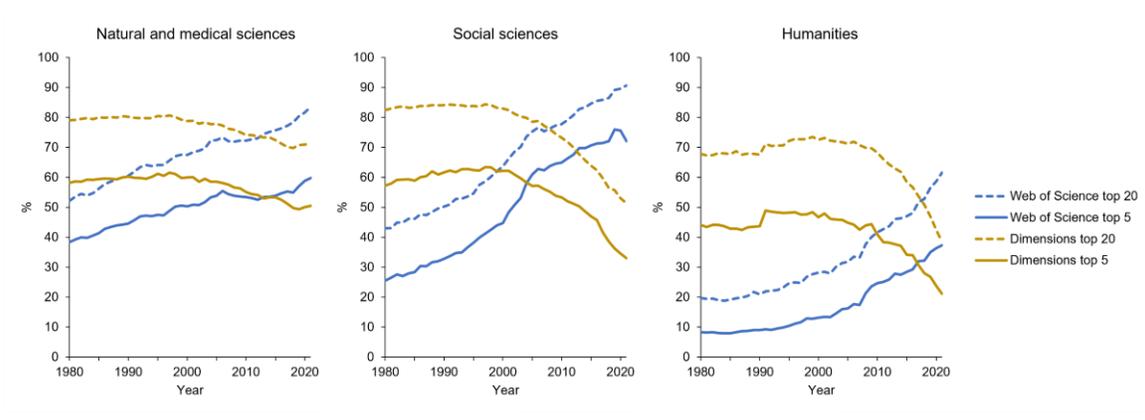

Figure 5: Proportions of articles published by the main publishers, per major discipline, according to two bibliometric databases, 1980-2021.

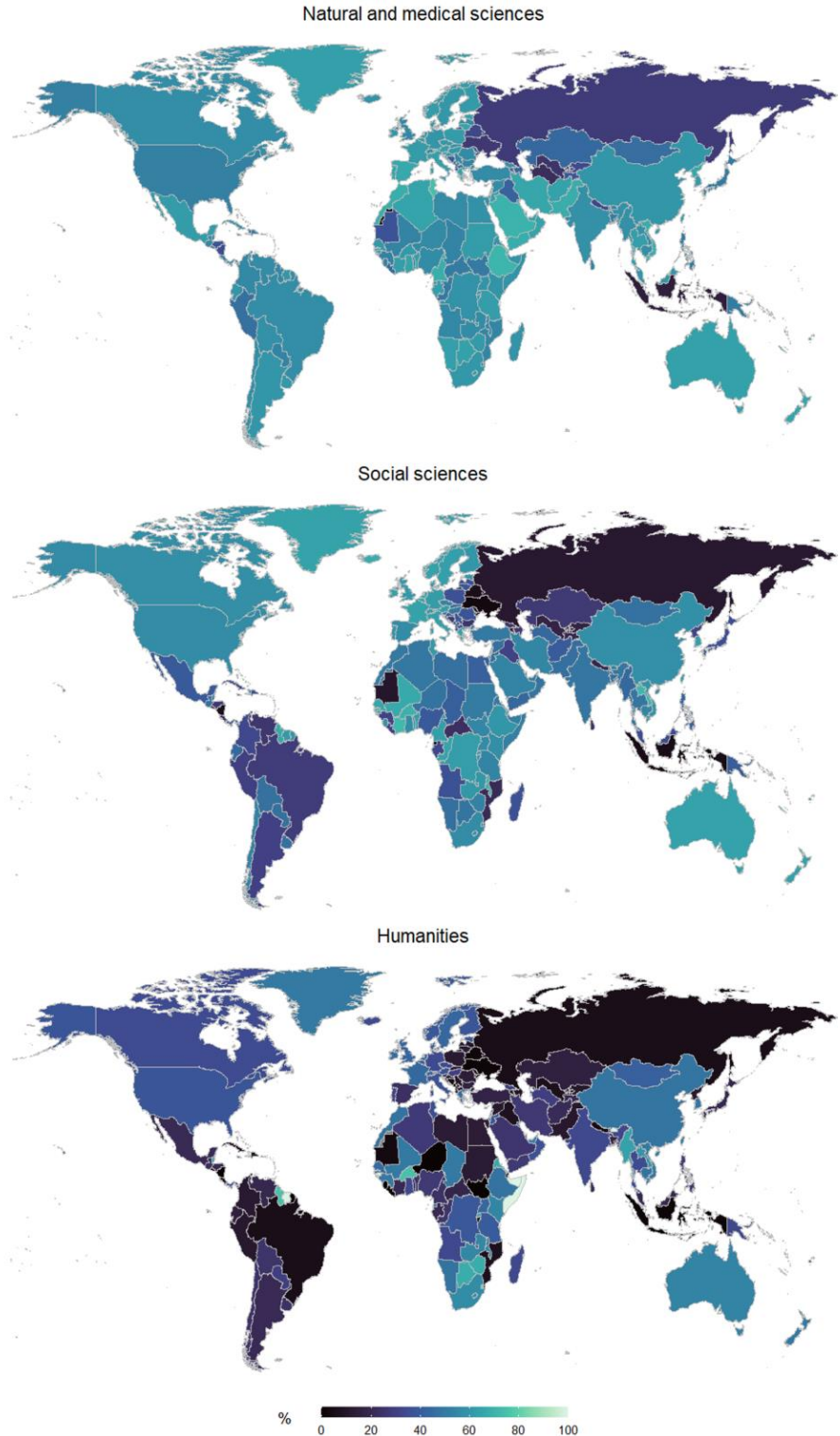

Figure 6: Proportion of articles published by the five major publishers (based on the number of articles), based on data from Dimensions (2019-2021). Articles were attributed to a country if at least one author originated from it.

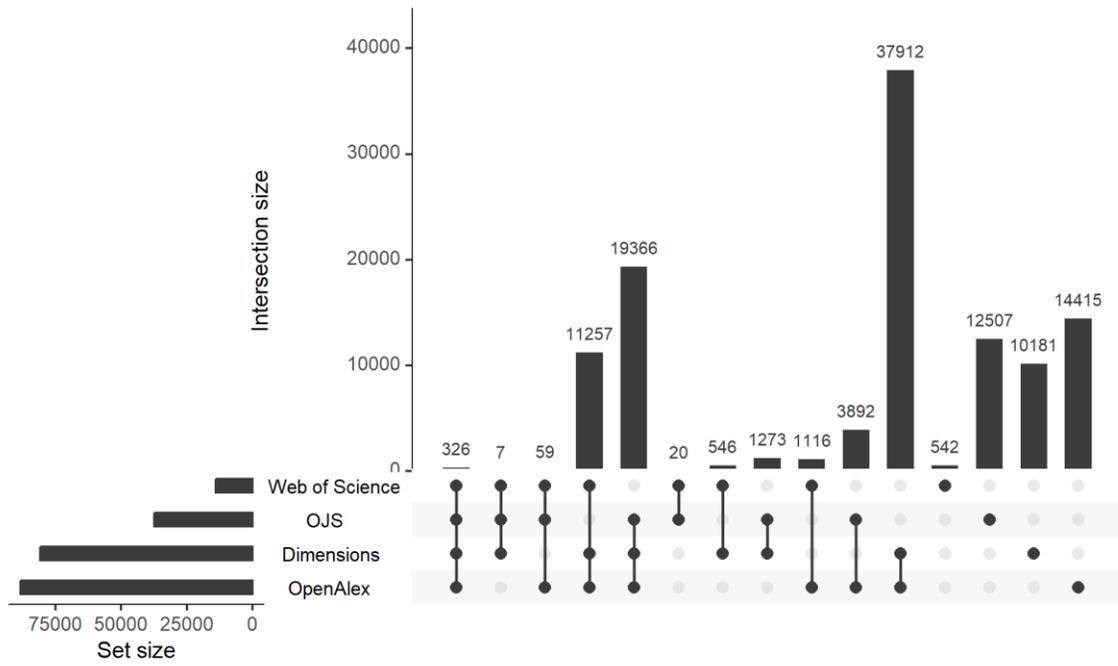

Figure 7: Comparison of the journal indexing of Web of Science, Dimensions and OpenAlex (2019-2021) and journals using OJS, including journals not using DOI for articles. The relationships (horizontal axis) are ordered by the degree of overlap between sets.